\documentclass[letterpaper,twocolumn,showpacs,prl]{revtex4} 
\usepackage{times}
\usepackage{graphics}
\usepackage{graphicx}
\usepackage{epsfig}
\usepackage{amsmath}
\begin{document}

\title{Phase diagram for a Bose-Einstein condensate moving in an optical lattice}
\author{Jongchul Mun}
\author{Patrick Medley}
\author{Gretchen K. Campbell}
\altaffiliation[current address: ]{JILA, Boulder, Colorado
80309, USA.}
\author{Luis G. Marcassa}
\altaffiliation[permanent address: ]{Instituto de Fisica de
S\~{a}o Carlos, University of S\~{a}o Paulo, S\~{a}o Carlos,
13560-970, SP, Brazil}
\author{David E. Pritchard}
\author{Wolfgang Ketterle}

\affiliation{ MIT-Harvard Center for Ultracold Atoms, Research
Laboratory of Electronics, and Department of Physics, MIT,
Cambridge, Massachusetts 02139, USA. }

\date{\today}

\begin{abstract}
The stability of superfluid currents in a system of ultracold bosons
was studied using a moving optical lattice. Superfluid currents  in
a very weak lattice become unstable when their momentum exceeds 0.5
recoil momentum.  Superfluidity vanishes already for zero momentum
as the lattice deep reaches the Mott insulator(MI) phase transition.
We study the phase diagram for the disappearance of superfluidity as
a function of  momentum and lattice depth between these two limits.
Our phase boundary extrapolates to the critical lattice depth for
the superfluid-to-MI transition with 2$\%$ precision. When a
one-dimensional gas was loaded into a moving optical lattice a
sudden broadening of the transition between stable and unstable
phases was observed.
\end{abstract}

\maketitle

The realization of condensed matter systems using ultracold atoms
brings the precision and control of atomic physics to the study of
many-body physics.  Many studies have focussed on Mott insulator
physics, an important paradigm for the suppression of transport by
particle correlations. Previous studies of the
superfluid(SF)-to-Mott insulator(MI) transition in optical lattices
with ultracold
bosons \cite{Jaksch-98,Orzel-01,Greiner-02,stoferle-04,Folling-05,Gerbier-05,Campbell-06,
Folling-06} adressed the quenching of superfluidity below a critical
lattice depth. Here we extend these studies into a second dimension
by studying stability of superfluid current as a function of
momentum and lattice depth as suggested in ref. \cite{Altman-05}.
These transport measurements show the stability of the quantum phase
in a moving system, which is far from equilibrium.

Transport measurements extend previous work on stationary systems in
two regards. First, superfluidity near the MI transition has only
been indirectly inferred from coherence measurements, whereas in
this work, we characterize the superfluid regime by observing a
critical current for superfluid flow. Second, previous
studies \cite{Jaksch-98,Orzel-01,Greiner-02,stoferle-04,Folling-05,Gerbier-05,Campbell-06,
Folling-06} were not able to precisely locate the phase transition,
since the observed excitation spectrum and atomic interference
pattern did not abruptly
change \cite{Greiner-02,Folling-05,Gerbier-05}, partially due to the
inhomogeneous density. In contrast, the sudden onset of dissipation
provides a clear distinction between the two quantum phases. In the
SF phase, current flows without dissipation if the momentum does not
exceed a critical momentum, while in the MI phase the critical
momentum vanishes and transport is dissipative.

Bosonic atoms in an optical lattice are often described by the
Bose-Hubbard Model where the tunneling between nearest neighbour
lattice sites is characterized by the hopping matrix element $J$ and
the repulsive interactions by the on-site matrix element
$U$ \cite{Jaksch-98,Fisher-89,Krauth-92,Freericks-94}.  The
dimensionless interaction energy $u\equiv U /J$ determines the
quantum phase of the system. For $u>u_c$, the system is in the MI
phase; for $u<u_c$, it is in the SF phase.

\begin{figure}
\centering{
\includegraphics[hiresbb=true,width=7 cm]{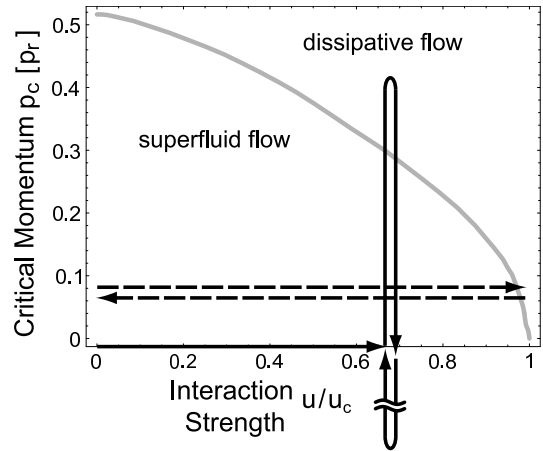}
\caption{Phase diagram showing the stability of superfluid flow in
an optical lattice and the experimental procedure. The grey curve
shows the predicted boundary between superfluid flow and dissipative
flow phases for a three-dimensional gas with a commensurate filling
of $N=1$ atom per site \cite{Altman-05}.  The solid (dashed) arrows
illustrate the experimental trajectory used for small (large)
lattice depths (see text for details).\label{expprocess} } }
\end{figure}

For weak interactions ($u \rightarrow 0$), the system approaches
single-particle physics in a periodic potential well described by
Bloch states and band structure.  The critical momentum for a stable
current-carrying state is $0.5$ $p_r$ ($p_r = h / \lambda$ is the
recoil momentum of an atom, where $\lambda$ is the wavelength of the
optical lattice light) \cite{Wu-2001}. At the critical momentum, it
becomes possible for two atoms in the same initial Bloch state to
scatter into two other states and conserve energy and
quasi-momentum \cite{Campbell-PRL06,Hilligsoe-05}. For sufficiently
deep lattices this occurs when the effective mass becomes negative.
Instabilities in a 1D optical lattice were studied theoretically
using a linear stability analysis of the Gross-Pitaevskii
Equation \cite{Wu-2001,Modugno-04} , and
experimentally \cite{Campbell-06,Fallani-04}. The theoretical studies
predicted that for increasing lattice depth or increasing atomic
interactions the stability of superfluid flow should
increase \cite{Wu-2001,Modugno-04}: the dynamic instability would
stay near $0.5$ $p_r$ whereas the Landau critical velocity and
therefore the energetic instability would shift to larger momenta
(For more discussions on dynamic and energetic instability, see
ref. \cite{Sarlo-05,BiaoWu-07}). However, these analyses neglect the
growing importance of quantum correlations for larger lattice depth
which lead to the SF-MI phase transition, where the critical
momentum for a superfluid current vanishes.  In this paper, we study
the decrease of the critical momentum from its value for the weakly
interacting regime towards zero at the MI transition.(Fig.
\ref{expprocess})

Most studies of the SF-MI phase transition monitor the
coherence in the superfluid phase through an interference
pattern observed in the ballistic expansion resulting from a
sudden turn-off of the confining potential and lattice.
Previous observations of the phase transition found the
experimental transition point to lie in the range between $10$
and $13$ $E_R$ (with the recoil energy defined as $E_R = p_r
^2 /2m$ where $m$ is the atomic mass) \cite{Greiner-02}.  This
uncertainty is related to the inhomogeneous density profile of
trapped atoms and to the fact that the visibility of the
interference extends beyond the transition point due to
short-range coherence in the MI phase \cite{Gerbier-05}.  It
has been suggested that observed kinks in the visibility are
linked to the formation of the MI shells with occupation
numbers $N=2$ and $3$ \cite{Gerbier-05}. Several authors have
suggested other features in the momentum distribution beyond
coherent interference peaks  as a more distinct signature of
the phase transition \cite{Wessel-04,Kashurnikov-02}.  Here we
show that the disappearance of the critical momentum for
superfluid flow provides such a signature and allows the
determination of the transition point with high precision.

Our measurement was not limited by the inhomogeneous density
profile.  For our range of lattice depths, low critical momenta and
the onset of dissipation occur only near the formation of MI shells
with integer occupation numbers $N$ \cite{Altman-05}. The onset of
dissipation related to the $N=1$ domains occurs at smaller momentum
than for other $N$ domains. For instance, with increasing momentum
$p$ the $N=1$ domain becomes unstable first, and this triggers
dissipation over the whole atomic cloud \cite{Altman-05}. Therefore,
the breakdown of superfluid flow in the system was determined by the
formation of the $N=1$ domain and was not smeared out by the
inhomogeneous density. Our criterion, the sudden onset of
dissipation, depended on the formation of an insulating shell
surrounded by a superfluid region, which occurs only in the
inhomogeneous case.

In our experimental setup, a Bose-Einstein condensate of $^{87}$Rb
atoms in the 5S$_{1/2}$ $\left|1,-1\right>$ state was prepared and
trapped in a combination of an Ioffe-Pritchard magnetic trap and an
optical dipole trap. The number of atoms in the BEC was typically $2
\times 10^5$, resulting in a maximum filling factor $N$ of around 3.
The magnetic trap frequencies were $\omega_{x,y}=40$ Hz radially and
$\omega_z=4.6$ Hz axially. The laser beam for the optical dipole
trap was oriented along the x-axis. This laser beam was
retroreflected and the polarization of the retroreflected beam was
rotated in order to minimize interference between the two beams.
Along the vertical direction (y-axis) a lattice was formed by a
retroreflected laser beam. For the z-axis, a moving lattice was
created by introducing a small frequency detuning $\delta f$ between
the two counter-propagating laser beams using acousto-optical
modulators driven by phase-locked frequency generators. The 3D
optical lattice was ramped up in the following way:  For the lattice
along the x-axis, the polarization of the retroreflected beam was
rotated to increase the interference contrast.  For the other two
axes, the power was increased exponentially in 160 ms. All lattice
beams were derived from the same laser operating at $\lambda=1064$
nm and had an $1/e^2$ waist of 100 $\sim$ 200 $\mu$m. The lattice
depth was calibrated with 1 $\%$ accuracy by applying a 12.5 $\mu$s
lattice laser pulse to a BEC and comparing the observed
Kapitza-Dirac diffraction pattern of a BEC to theory. Fits of more
than twenty diffraction patterns for the lattice depth from $<$1
$E_R$ to 40 $E_R$ resulted in an accuracy of the calibration of 1
$\%$.


\begin{figure}
\centering{
\includegraphics[hiresbb=true,width=7 cm]{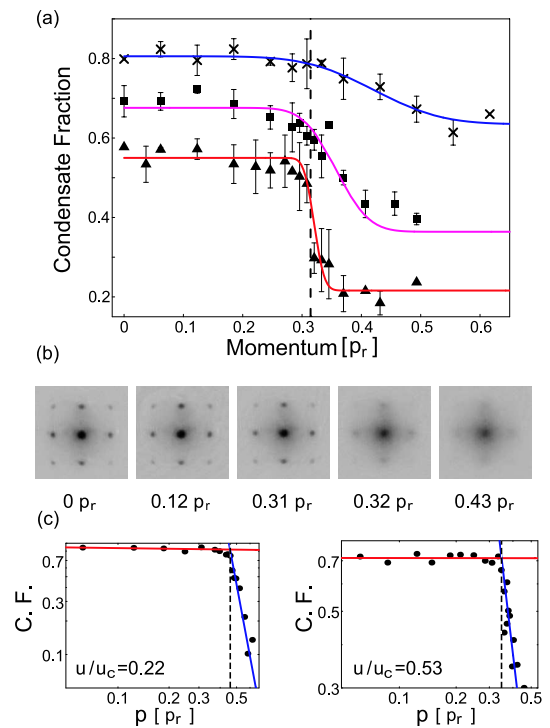}
\caption{Determination of the critical momentum of superfluid flow.
Shown is the condensate fraction as a function of a momentum $p$.
(a) Condensate fraction with $u/u_c=0.61$ for a variable number of
cycles of the momentum modulation (one cycle: cross marks and blue
line, two cycles:  squares and purple line, three cycles: triangles
and red line). A dashed vertical line indicates the critical
momentum where instability begins to occur. The two and three-cycle
data are offset vertically for clarity. These data were fitted with
an error function to guide the eye. (b) Images of interference
patterns released from an optical lattice at $u/u_c=0.61$ moving
with variable momentum. Instability occurred between $p=0.31p_r$ and
$0.32p_r$. Some of the triangular data points in (a) were obtained
from these images. (c) Condensate fraction on a log-log scale for
two different interaction strengths. \label{CFplot} } }
\end{figure}

For transport measurements, we moved an optical
lattice \cite{Fallani-04,Dahan-96} which provides more flexibility to
change the momentum than exciting a dipole oscillation by
displacement of the BEC \cite{Cataliotti-01,Fertig-05}. A moving
optical lattice with velocity $v=\lambda \cdot \delta f/2$ was
created along the long axis of the BEC by introducing a small
frequency detuning $\delta f$ between two counter-propagating
lattice beams. If the velocity $v(t)$ changes slowly enough not to
induce interband excitations, the initial Bloch state
$\left|p=0\right>$  of the condensate in the optical lattice
adiabatically evolves into the current-carrying state $\left|
p(t)=-mv(t) \right>$ where $p$ is the quasi-momentum. For increasing
lattice depth, the effective mass of atoms $m^*=[\partial^2
E(p)/\partial p^2]^{-1}$ increases, and the group velocity $v_g =
-(m/m^*) v(t)$ decreases. As a consequence, atoms prepared in a
moving lattice with quasi-momentum $p = -m v$ travel in the frame of
the moving lattice with $v_g$ and in the lab frame with velocity
$\Delta v = v+v_g =(1-m/m^*)v$, which approaches $v$ in a deep
lattice. Consequently, we observed that in a deep moving lattice
atoms were dragged along to the edge of the trapping region limiting
the experimental time scale to probe for dissipation.  This became
an issue for larger values of $p$ and was addressed by first ramping
up the lattice with $p=0$ and then alternating the velocity of the
moving lattice, thus performing a low-frequency AC transport
measurement instead of DC.

We have used two sets of experimental procedures (Fig.
\ref{expprocess}), and our results were consistent for both. Close
to the SF-MI phase transition, the lattice was increased to
$V_{latt}$ with a fixed (and small) value of momentum $p$ (dahsed
arrows in Fig. \ref{expprocess}). After a variable hold time
$t_{hold}$  at $V_{latt}$ the lattice was ramped down to zero, and
the magnetic trap switched off. After 33 ms of ballistic expansion,
the atoms were imaged and the condensate fraction was determined as
a function of momentum. For smaller lattice depths, the lattice was
ramped up with $p=0$ (Fig.\ref{expprocess}). Then a sinusoidal
momentum modulation of the moving lattice with amplitude $p_M$ was
applied by modulating the frequency detuning $\delta f$ between the
counter-propagating lattice beams. The 10 ms period of this momentum
modulation was slow enough to meet the adiabaticity condition, but
fast enough to limit the displacement of the atomic cloud to less
than a few $\mu$m. Both the trapping potential and the optical
lattice were then turned off suddenly. After 33 ms of ballistic
expansion, the condensate fraction of the center peak of the
superfluid interference pattern was recorded as a function of the
momentum modulation amplitude $p_M$.  Several cycles (typically,
three to five) of the momentum modulation were applied to obtain a
high contrast between the stable and dissipative regimes (Fig.
\ref{CFplot} (a)).

Fig. \ref{CFplot} (a) shows how the transition between superfluid
and dissipative currents became sharper with increasing number of
cycles of the momentum modulation. The critical momentum was
determined from a log-log plot of the condensate fraction as a
function of momentum $p$ (Fig. \ref{CFplot} (c)). For low $p$, the
condensate fraction is constant. For high $p$, the condensate
fraction decreases. The intersection between two linear fit
functions was taken as the critical momentum. Our result was found
to be independent of the time period and number of cycles of the momentum
modulation at a few percent level.

\begin{figure}
\centering{
\includegraphics[hiresbb=true,width=7 cm]{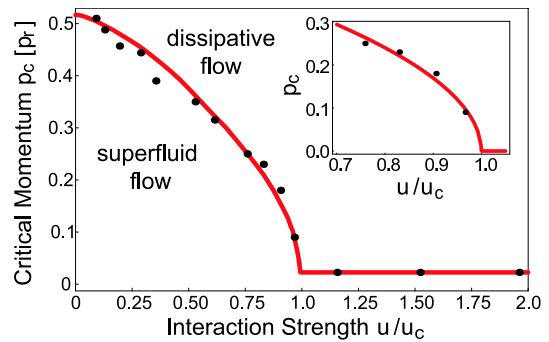}
\caption{ Critical momentum for a condensate in a 3D lattice. The
solid line shows the theoretical prediction for the superfluid
region.  The horizontal solid line is a fit to the data points in
the MI phase. (Inset) Fit of critical momenta near the SF-MI phase
transition. \label{phasediagram} } }
\end{figure}

In the MI phase, stable superfluid flow is not possible and the critical
momentum should vanish.  However, using the procedure described
above, we measured a small critical momentum of $0.02$ $p_r$ for
lattice depths $V_{latt}=14, 15, 16$ $E_R$. Up to this momentum, the
SF-MI phase transition remained reversible. We interpret the
non-zero critical momentum as a finite-size effect.  For our cloud
size of $60$ $\mu m$, the corresponding Heisenberg momentum
uncertainty of $0.018$ $p_r$ agrees with our measured critical
momentum.  In cold atom experiments, some sloshing motion of the
cloud in the trapping potential is unavoidable.  The momentum
uncertainty determined above indicates how much sloshing motion can
be tolerated without affecting the observed phase transition.

The critical lattice depth for the SF-MI phase transition can be
determined as the point where the critical momentum vanishes.  Using
the predicted functional form \cite{Altman-05} of the approach
towards zero, $p_c \propto \sqrt{1-u/u_c}$, as a fit function for
the data points close to the SF-MI phase transition (the data points
shown in the inset of Fig. \ref{phasediagram}) we determined the
critical value $u_c = 34.2$ $(\pm 2.0)$ corresponding to a lattice
depth of $13.5(\pm0.2)$ $E_R$ \footnote{The lattice depth was converted to dimensionless
interaction energy $u$ following the method in ref. \cite{Jaksch-98} using Wannier function truncated
at the 5th excited Bloch band and a s-wave scattering length
$a=100.44 a_0$ \cite{harber-02} ($a_0$: Bohr radius). }. Our result
agrees with the mean field theory prediction $u_c =5.8\times6=34.8$
for $N=1$ SF-MI phase transition \cite{Jaksch-98} and deviates by 2
$\sigma$ from the predictions of $u_c=29.34(2)$ of quantum Monte
Carlo(QMC) simulation \cite{Freericks-96,capogrossosansone-07}, which
includes corrections beyond the mean field theory. This demonstrates
that our method has the precision to identify non-mean field
corrections. However, to turn precision into accuracy, experiments
or QMC simulations \cite{Wessel-04,Freericks-96,capogrossosansone-07}
have to address corrections due to finite size, finite temperature,
and finite time to probe the onset of the
instability \cite{capogrossosansone-07}. In our experiment, these
corrections seemed to be small, but have not been characterized at
the level of 1$\%$ in lattice depth.

\begin{figure}
\centering{
\includegraphics[hiresbb=true,width=7 cm]{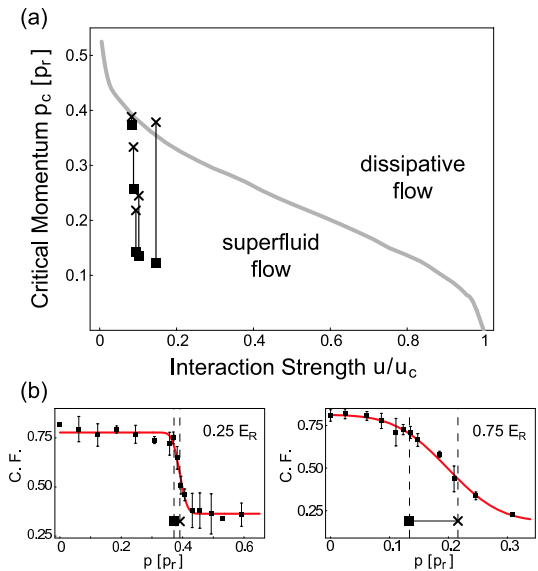}
\caption{ Critical momentum for a 1D gas in an optical
lattice. (a) The grey line indicates the mean-field theory
prediction.  The interaction strengths are normalized by the
mean-field prediction for $u_c=5.8 \times 2$
 \cite{Jaksch-98,stoferle-04}. Squares(crosses) represent the
measured critical momentum (the center of the transition).
Measurements were taken at lattice depths of $0.25, 0.50,
0.75, 1.0, 2.0$ $E_R$. The lines between crosses and squares
indicate the width of the transition region.  (b) Condensate
fraction measured at $0.25$ $E_R$ (left) and $0.75$ $E_R$
(right).The data were fitted with an error function. Squares
(the critical momentum) and crosses (the center of the
transition) are indicated on the plots. \label{1ddata} } }
\end{figure}

The mean-field prediction for stable superfluid flow in 1D is
similar to that for the 3D system \cite{Altman-05}.  However,
it is well known that fluctuations play a much more important
role in 1D. For studying a 1D system, we prepared an array of
one-dimensional gas tubes by ramping two pairs of optical
lattice beams up to lattice depths of $V_x=V_y=30$ $E_R$
suppressing hopping between the tubes.  After a hold time of
10 ms, a moving optical lattice was ramped up along the
z-axis.  As in our 3D experiment, a momentum modulation was
applied, after which the moving optical lattice was ramped
down to zero, followed by the other two optical lattices.  The
condensate fraction was determined after 33 ms of ballistic
expansion as a function of the momentum modulation amplitude.
The critical momentum, where the onset of dissipation begins,
was identified from a log-log plot as in the 3D case. Since
the transitions became very broad, we characterized them by an
error function fit, with the center of the fitted error
function taken as the center of the transition (Fig.
\ref{1ddata}).

In the 1D system, at a very shallow lattice depth of $0.25$ $E_R$
(corresponding to $u/u_c= 0.08$) a sharp transition was observed,
and the measured critical momentum agreed very well with the
prediction \cite{Altman-05,Polkovnikov-05} of a critical momentum of
$0.39$ $p_r$. However, a slight increase of the interaction strength
(to $u/u_c=0.09$ at a lattice depth of $0.5$ $E_R$) led to a
significant decrease of the critical momentum as well as a dramatic
broadening of the transition as shown in Fig. \ref{1ddata}. For
lattice depths larger than $2$ $E_R$, the transition became very
broad and showed complex behavior, and we could not obtain
quantitative fits.  Our results show a significant deviation from
the mean-field theory predictions and are in agreement with previous
experiments \cite{Fertig-05} where damped dipole oscillations of a 1D
Bose gas in an optical lattice were observed, and the damping grew
rapidly even at very shallow lattice depths around $0.25$ $E_R$.

The observed broadening of the transition confirms theoretical
studies which emphasize the importance of quantum fluctuations in
the 1D system.  Quantum tunneling out of metastable states which are
ignored in the mean-field description can lead to a decay of the
superfluid current at very low momenta \cite{Polkovnikov-05}.  In
addition to quantum fluctuations, thermal fluctuations provide a
mechanism for current decay \cite{Polkovnikov-05}.  In our
experiment, we used a ``pure'' BEC without a discernible thermal
component. The close agreement with $T=0$ predictions indicates that
thermal fluctuations were not dominant.


In conclusion we have used transport studies to connect a well-known
dynamical instability for weakly interacting bosons with the
equilibrium superfluid to Mott insulator transition.  A
comparison of 3D and 1D systems confirms the applicability of a
mean-field description in 3D and the crucial importance of
fluctuations in 1D.  The disappearance of superfluid currents at the
SF-MI phase transition precisely located the phase transition.
Our results illustrate the control and precision of condensed matter
physics experiments done with ultracold atoms and their suitability
to test many-body theories.

\begin{acknowledgments}
This work was funded by NSF through the grant for CUA. L.G.M.
acknowledges support from Coordenacao de Aperfeicoamento de
Pessoal de Nivel Superior. We thank E. Demler and A.
Polkovnikov for insightful discussions, and David Weld for a
critical reading of the manuscript.
\end{acknowledgments}

\end{document}